\documentclass[11pt]{article}

\usepackage{acl}

\usepackage{times}
\usepackage{latexsym}
\usepackage{amsmath}
\usepackage{amssymb}
\usepackage{tikz}
\usepackage{listings}
\usepackage{xcolor}
\usepackage{tcolorbox}
\usepackage{makecell} 
\usepackage{booktabs}
\tcbuselibrary{listings,skins,breakable}

\newcommand{\tool}{{BanglaForge}~}
\newcommand{\toolnospace}{BanglaForge}
\usepackage[T1]{fontenc}

\usepackage[utf8]{inputenc}

\usepackage{microtype}
\usepackage{algorithm}
\usepackage{algpseudocode}
\usepackage{enumitem}
\usepackage{inconsolata}

\usepackage{graphicx}

\definecolor{codegreen}{rgb}{0,0.6,0}
\definecolor{codegray}{rgb}{0.5,0.5,0.5}
\definecolor{codepurple}{rgb}{0.58,0,0.82}
\definecolor{backcolour}{rgb}{0.95,0.95,0.92}

\lstdefinestyle{mystyle}{
    backgroundcolor=\color{backcolour},   
    commentstyle=\color{codegreen},
    keywordstyle=\color{magenta},
    numberstyle=\tiny\color{codegray},
    stringstyle=\color{codepurple},
    basicstyle=\ttfamily\footnotesize,
    breakatwhitespace=false,         
    breaklines=true,                 
    captionpos=b,                    
    keepspaces=true,                 
    numbers=left,                    
    numbersep=5pt,                  
    showspaces=false,                
    showstringspaces=false,
    showtabs=false,                  
    tabsize=2
}

\title{\toolnospace: LLM Collaboration with Self-Refinement for Bangla Code Generation}

\author{
Mahir Labib Dihan, Sadif Ahmed, Md Nafiu Rahman \\
Department of Computer Science and Engineering \\
Bangladesh University of Engineering and Technology (BUET) \\
Dhaka, Bangladesh \\
\texttt{\{mahirlabibdihan, ahmedsadif67, nafiu.rahman\}@gmail.com}
}

\begin{document}
\maketitle
\begin{abstract}
Bangla is a low-resource language for code generation, lacking large-scale annotated datasets and tools to transform natural language specifications into executable programs. This makes Bangla-to-code generation a challenging task requiring innovative solutions. To address this, we introduce \textbf{\toolnospace}, a novel framework for generating code from Bangla function descriptions. 
\tool leverages a retrieval-augmented dual-model collaboration paradigm with self-refinement, combining in-context learning, llm-based translation, systematic prompt engineering, and iterative self-refinement based on execution feedback, where a coder generates initial solutions and a reviewer enhances them for robustness. On the \textbf{BLP-2025 Bangla Code Generation} benchmark, \tool achieves a competitive Pass@1 accuracy of \textbf{84.00\%}, demonstrating the effectiveness of retrieval, model collaboration, and self-refinement for low-resource Bangla code generation.
\end{abstract}

\section{Introduction}
Large language models (LLMs) have shown strong capabilities in code generation, where natural language descriptions are automatically transformed into executable programs. Models such as Codex, CodeT5, and StarCoder, trained on large-scale code–text corpora, can produce syntactically valid and semantically correct solutions, performing well on benchmarks like HumanEval \cite{chen2021evaluating}. These advances reduce the gap between human intent and code, making programming more accessible. However, most existing systems are designed for English inputs, leaving low-resource languages underserved. Models often struggle with informal structures, domain-specific terms, and semantic nuances, resulting in incorrect or brittle outputs.

\noindent We introduce \textbf{\toolnospace}, a framework for generating executable code from Bangla task descriptions. Each input is represented as a triple: the Bangla description, its English translation, and unit test assertions. This structure leverages the model’s stronger English understanding while retaining Bangla context. \tool combines retrieval-augmented prompting, iterative self-refinement with execution feedback, and a dual-model coder–reviewer pipeline. Our system achieves a \textbf{Pass@1 accuracy of 84\%} on \textbf{BLP-2025 Bangla Code Generation Benchmark}~\cite{raihan-etal-2025-blp}, demonstrating the potential of practical low-resource code generation.

\noindent Our contributions can be summarized as follows:
\begin{itemize}[itemsep=1pt,topsep=0pt, leftmargin=*]
    \item A retrieval-augmented few-shot prompting approach using TF-IDF to select relevant Bangla--Python pairs, improving in-context learning despite limited labeled data.
    \item A LLM-based translation component that translates Bangla instructions into English with the help of a glossary to enable accurate cross-lingual code generation.
    \item An iterative self-refinement protocol that leverages execution feedback to detect and correct errors across refinement cycles.
    \item A dual-model architecture where a generator model focuses on functional correctness and a reviewer model enhances robustness, style, and coverage of edge cases.
\end{itemize}

\noindent We release our implementation of \tool at \url{https://github.com/mahirlabibdihan/BanglaForge} to facilitate reproducibility and further research.

\begin{figure*}[t]
  \centering
  \includegraphics[width=0.96\linewidth]{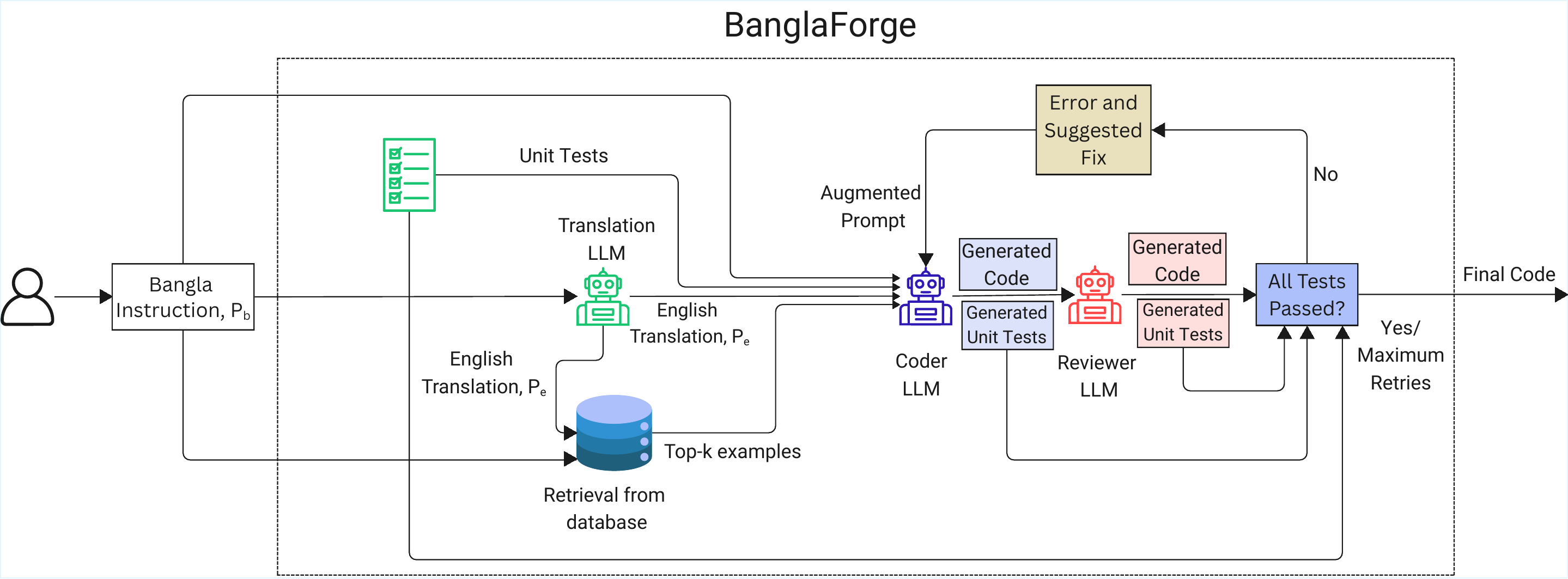}
  \caption{Workflow of the proposed \textbf{\tool} framework. 
A Bangla instruction ($P_b$) is translated into English ($P_e$) and, together with unit tests, used to retrieve top-$k$ bilingual examples. 
The \textbf{Coder LLM} then generates Python code and additional test cases. 
The \textbf{Reviewer LLM} validates, refines, and re-prompts upon errors until all tests (original and generated) are passed, yielding the final code.}
\label{fig:blp-workflow}
\end{figure*}

\section{Related Works}

Research in Bangla NLP has evolved from early word embeddings to specialized LLMs. Initial efforts such as BnVec introduced embeddings like fastText, Word2Vec, and GloVe trained on diverse corpora, with customized fastText outperforming multilingual baselines in classification tasks \cite{kowsher2021bnvec,enhanced_fasttext_transfer2022,mojumder2020fasttext}. Recent advances include Bangla LLMs and benchmarks such as TigerCoder \cite{raihan2025tigercoder} and BanglaByT5 \cite{banglabyT52025}, which advanced code generation and tokenization strategies. However, existing work largely focuses on pretraining and benchmarking without complete generation pipelines. Our work addresses these gaps by introducing retrieval-augmented prompting, iterative self-refinement, and a dual generator–reviewer design. A detailed discussion is provided in Appendix \ref{sec:appendix_rw}.

\section{Dataset}

We build on the resources introduced for Bangla code generation across recent shared tasks and benchmarks. Our dataset comes from the Bangla Code Generation shared task (Task 2) at BLP-2025~\cite{raihan-etal-2025-blp}, where the objective is to translate Bangla natural language programming prompts into Python functions that satisfy hidden unit tests. The dataset is distributed through an official starter kit\footnote{\url{https://noshinulfat.github.io/blp25_code_generation_task/\#/get-started}}, which also provides baseline code and evaluation scripts.  

Each entry is a JSON object containing four fields: an \texttt{id}, a Bangla instruction describing the task, a \texttt{response} field with the reference Python implementation (training only), and a \texttt{test\_list} field of assert-based unit tests. 

\begin{figure}
    \centering
    \includegraphics[width=\linewidth]{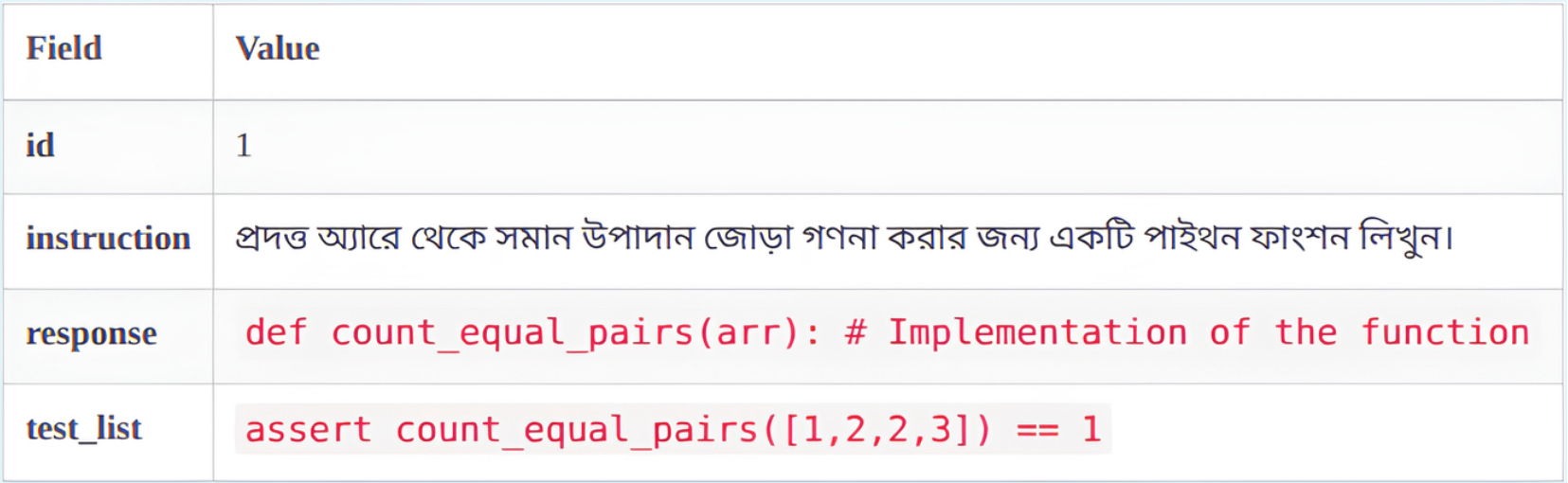}
    \caption{Example data point}
    \label{fig:data-point}
\end{figure}

\begin{table}[h]
\centering
\label{tab:dataset-split}
\begin{tabular}{|l|c|c|}
\hline
\textbf{Split} & \textbf{Purpose} & \textbf{Size} \\ 
\hline
Trial & Initial experiments & 74 \\ 
Development & Validation & 400 \\ 
Test & Final evaluation & 500 \\ 
\hline
\end{tabular}
\caption{Dataset Split Statistics for Bangla Code Generation}
\end{table}

For development and testing, we adopt two external Bangla code generation benchmarks. The \textbf{mHumanEval-Bangla} dataset~\cite{raihan-etal-2025-mhumaneval}, a Bangla extension of HumanEval, is used during the development phase, enabling programmatically testable evaluation on held-out prompts. The \textbf{MBPP-Bangla} dataset~\cite{raihan2025tigercoder}, adapted from MBPP as part of the TigerCoder framework, is used during both development and test phases, providing diverse programming problems in Bangla with associated unit tests. 

\section{Methodology}

We propose \textbf{\toolnospace}, a retrieval-augmented dual-LLM framework for generating Python code from Bangla natural language specifications. The system tackles low-resource code generation through structured prompt design, bilingual translation, example retrieval, and a two-stage generation-review process involving a \textit{Coder LLM} and a \textit{Reviewer LLM}. Together, these components ensure both functional correctness and stylistic reliability, even in underrepresented languages like Bangla. An overview of the complete workflow is shown in Figure~\ref{fig:blp-workflow} and in Algorithm~\ref{alg:banglaforge} (Appendix). Each stage is described in detail below.

\subsection{Problem Formulation and Input Representation}

Each task in the dataset consists of a Bangla instruction $P_b$ and its corresponding public unit tests $T=\{t_1,\dots,t_n\}$. To enable code synthesis, the instruction is translated into English using a translation model equipped with a controlled glossary for mathematical and algorithmic terms (e.g., GCD,  LCM, sum). The glossary is curated by the authors which was motivated from the provided dataset and commonly seen technical terms in code related works. The translated instruction $P_e$ retains the semantic fidelity of the Bangla instruction while ensuring syntactic clarity for code generation. The system’s objective is to synthesize a Python function $f$ such that all $t_i \in T$ are satisfied given the constraints in $P_b$. Function prototypes are normalized to valid Python syntax, aligning argument and return types with unit test definitions.

\subsection{Retrieval-Augmented Example Selection.}

To enhance contextual understanding, both Bangla and English task descriptions are used to retrieve semantically similar solved examples from a bilingual database $\mathcal{D} = \{(p_i^b, p_i^e, c_i, T_i)\}_{i=1}^{N}$. Each entry contains the Bangla and English prompts, the reference code ($c_i$), and associated test cases ($T_i$). Both $P_b$ and $P_e$ are embedded using TF-IDF unigram bigram representations. We chose TF-IDF due to its high computational efficiency and strong performance on smaller datasets, as dense retrievers typically require a large training corpus to be effective \cite{arabzadeh2021predicting}. For our task, TF-IDF's strength in matching exact, high-signal technical keywords (e.g., GCD,'' factorial'') is paramount. This lexical precision provides a fast and more reliable baseline for retrieving analogous code problems than a dense model's generalized semantic understanding \cite{karpukhin2020dense}.
The top-$k$ examples (typically $k=5$) are selected and inserted into the prompt as few-shot exemplars. 
For experiments on the Development set, the database $\mathcal{D}$ consists of the Trial set. For experiments on the Test set, we use the combined Trial+Development sets as the database.
This bilingual, retrieval-augmented setup enables contextual grounding and helps the model capture problem-solving patterns from similar tasks. The retrieved example format is provided in Appendix \ref{sec:appendix_prompts}.

\subsection{Stage 1: Code Generation by Coder LLM.}

The \textit{Coder} LLM receives a composite input consisting of the Bangla instruction $P_b$, English translation $P_e$, the retrieved top-$k$ example pairs $(p_i^b, p_i^e, c_i, T_i)$, and the provided unit tests $T$. Based on this augmented prompt, the Coder LLM generates a Python code candidate $c_0$ intended to satisfy $T$, and
additional synthetic test cases $T_c$ designed to cover potential edge or missing cases. This stage focuses on functional code generation guided by contextual analogies from retrieved examples. The output $(c_0, T_c)$ is then passed to the Reviewer LLM for refinement. The detailed prompt for Coder LLM is provided in Appendix \ref{sec:appendix_prompts}.

\begin{table*}[t]
\centering
\small
\renewcommand{\arraystretch}{1.05}
\begin{tabular}{l|c|c|c|c|c}
\hline
\textbf{Model} &
\makecell{\textbf{Few Shot}} &
\makecell{\textbf{\# Examples}} &
\makecell{\textbf{Translation}} &
\makecell{\textbf{\# Unit Tests}} &
\textbf{Pass@1} \\
\hline
\multicolumn{6}{c}{\textbf{Dev Set}} \\
\hline
Gemma-1B & N/A & 0 & No & 0 & 27.25\% \\
GPT-OSS-20B & Manual & 3 & No & 0 & 60.25\% \\
GPT-OSS-20B & Manual & 5 & No & 0 & 61.25\% \\
DeepSeek-R1-Llama-70B & Manual & 5 & Yes & 0 & 57.75\% \\
Gemini-2.0-Flash & Manual & 3 & Yes & 0 & 60.00\% \\
Gemini-2.0-Flash & Manual & 5 & Yes & 0 & 62.50\% \\
Lg Exaone Deep 32B & Manual & 5 & Yes & 1 & 85.25\% \\
Lg Exaone Deep 32B & Manual & 5 & Yes & 3 & 94.25\% \\
Lg Exaone Deep 32B & RAG (Trial) & 5 & Yes & 3 & 95.50\% \\
\hline
\multicolumn{6}{c}{\textbf{Test Set}} \\
\hline
Lg Exaone Deep 32B & RAG (Trial+Dev) & 5 & Yes & 1 & 80.60\% \\
\textbf{Gemini-2.5-Pro} & \textbf{RAG (Trial+Dev)} & \textbf{5} & \textbf{Yes} & \textbf{1} & \textbf{84.00\%} \\
\hline
\end{tabular}
\caption{Pass@1 accuracy of models on the BLP-2025 Development and Test sets.}
\label{tab:results}
\end{table*}

\subsection{Stage 2: Code Review and Refinement by Reviewer LLM.}

The \textit{Reviewer} LLM acts as a validator and refiner. It takes as input the code and test cases generated by the Coder LLM along with the original task description and unit tests. Its responsibilities include running static and logical checks on $c_0$, correcting syntax or runtime issues, improving variable naming, structure, and input validation, generating an additional set of refined unit tests $T_r$ to ensure covering edge cases. If any error or inconsistency is detected, the Reviewer LLM suggests an explicit fix and the process is repeated up to a maximum of $M$ iterations ($M=5$). The detailed prompt for Reviwer LLM is provided in Appendix \ref{sec:appendix_prompts}.

\subsection{Iterative Self-Refinement Protocol.}

The refinement loop is formally defined as: $c_{i+1} = \mathcal{R}(c_i, e_i, \mathcal{P})$,
where $\mathcal{R}$ denotes the Reviewer LLM, $e_i$ is the detected error, and $\mathcal{P}$ represents the augmented prompt containing feedback. The cycle continues until all test cases—original ($T$), coder-generated ($T_c$), and reviewer-generated ($T_r$)—are successfully passed, or until the retry limit $M$ is reached. This multi-level testing ensures that the final solution generalizes beyond the given test cases. The errors and suggested fixes are provided in Appendix \ref{sec:appendix_prompts}.

\section{Experiment}
\subsection{Evaluation Metrics} 
We evaluate performance using the Pass@1 accuracy metric, which measures the proportion of problems solved correctly in the first iteration. This metric provides a clear and direct assessment of the system’s accuracy in solving problems without requiring further refinements.

\subsection{Models}

We evaluate several large language models (LLMs) for Bangla code generation. 
The models tested on the Development set include \textbf{Gemma-1B}~\cite{gemma2024}, 
\textbf{GPT-OSS-20B}~\cite{gptoss2024}, 
\textbf{DeepSeek-R1-Llama-70B}~\cite{deepseek2025}, 
\textbf{Gemini-2.0-Flash}~\cite{gemini2024}, 
and \textbf{Lg Exaone Deep 32B}~\cite{exaone2024}, 
with different prompting strategies and unit-test settings. 
For the final evaluation on the Test set, we select \textbf{Lg Exaone Deep 32B}~\cite{exaone2024} 
and \textbf{Gemini-2.5-Pro}~\cite{gemini2025} under their best-performing configurations within a retrieval-augmented dual-stage pipeline.

\subsection{Results}

We evaluate our system on the BLP-2025 Bangla code generation benchmark. 
Our experiments are conducted in two stages: first on the Development set to explore different models and prompting strategies, and then on the Test set to report final results. 
Table~\ref{tab:results} presents the Pass@1 accuracy for various models and configurations across both sets.

The development set results reveals that small-scale models such as \textbf{Gemma-1B} achieve only 27.25\% Pass@1, underscoring the challenge of Bangla-to-code translation without contextual guidance. Larger open-source models like \textbf{GPT-OSS-20B} shows improvements (60.25-61.25\%) under few-shot prompting, though performance gains taper off with additional in-context examples. Introducing translation-based prompting further improves comprehension of Bangla instructions, as seen with \textbf{DeepSeek-R1-Llama-70B} (57.75\%) and \textbf{Gemini-2.0-Flash} (60-62.5\%).  

A major performance leap is observed with the \textbf{Lg Exaone Deep 32B} model, which combines translation and lightweight unit-test feedback. Accuracy rises from 85.25\% with one visible test to 94.25\% with three tests, highlighting the benefit of guided reasoning through intermediate validation. When enhanced with our RAG pipeline on the trial set, the model achieves 95.5\% Pass@1 on the development benchmark—demonstrating consistent improvements through contextual retrieval and refinement.

On the held-out test set, the RAG-augmented \textbf{Lg Exaone Deep 32B} achieves 80.6\% Pass@1, while the more recent \textbf{Gemini-2.5-Pro} model further pushes performance to \textbf{84.0\%}. These results confirm that retrieval augmentation combined with multilingual comprehension yields robust generalization across unseen Bangla programming tasks.


\section{Conclusion}

In this paper, we presented a retrieval-augmented dual-model framework for generating Python code from Bangla instructions. Combining structured prompting, iterative self-refinement, and a generator-reviewer design, our system achieved Pass@1 accuracy of 84\% on the BLP-2025 benchmark. The approach consistently outperforms baselines, showing the effectiveness of retrieval augmentation and feedback-driven refinement for low-resource code generation. Future work will expand the framework to other languages and incorporate reinforcement-based refinement. Additionally, improvements in RAG corpus and Bangla-to-English translation quality are expected to further enhance the overall performance of the pipeline.

\section{Limitations}

While BanglaForge demonstrates strong performance on the BLP-2025 Bangla code generation benchmark, several limitations remain.
First, the system relies heavily on high-quality bilingual translation; inaccuracies in Bangla-to-English mapping or glossary coverage can propagate errors to the generation stage. Second, the retrieval component depends on TF-IDF, which captures lexical overlap but may miss deeper semantic similarities, especially in complex algorithmic prompts. Third, the framework assumes well-structured Bangla input; informal phrasing or dialectal variations could reduce translation fidelity and retrieval relevance. Additionally, self-refinement cycles are limited to a fixed number of iterations and do not incorporate adaptive stopping or learning from prior refinements.
Finally, since the dataset itself originates from machine-translated English sources, true Bangla-native problem framing and linguistic diversity remain under-represented. Future work should explore human-curated datasets, semantic retrieval models, and reinforcement-based refinement to address these limitations.
\bibliography{custom}
\appendix
\section{Related Works}
\label{sec:appendix_rw}
The trajectory of research in Bangla NLP has shifted from foundational embeddings and lightweight classification models to full-fledged Bangla LLMs and, more recently, toward modular architectures that integrate retrieval and feedback. In the early days, emphasis was placed on crafting vector representations tailored to Bangla’s morphological richness and vocabulary distribution. The BnVec project, for instance, introduced Bangla-specific fastText, Word2Vec, and GloVe embeddings that placed importance on vocabulary coverage and representation quality \cite{kowsher2021bnvec}. Later work showed that embeddings trained on Bangla corpora outperform multilingual embedding baselines in text classification and related tasks \cite{enhanced_fasttext_transfer2022, mojumder2020fasttext}. Meanwhile, the Vacaspati corpus and derived models such as Vac-FT and Vac-BERT demonstrated that diversifying corpus domains and scaling data can boost embedding and language model utility beyond standard fastText baselines \cite{vacaspati2023}.

As the field progressed, researchers began developing Bangla-centric pretrained language models for both understanding and generation. A notable early example is BanglaBERT, introduced by Bhattacharjee et al., which is a BERT (ELECTRA-discriminator)–style model pretrained on a 27.5 GB Bangla corpus (“Bangla2B+”) and evaluated on a suite of Bangla NLU benchmarks that include classification, NLI, NER, and QA tasks under the BLUB benchmark \cite{bhattacharjee-etal-2022-banglabert}. BanglaBERT outperforms multilingual baselines on those tasks, showing that language-specific pretraining brings tangible gains in low-resource settings. Building on that, more recent works such as enhanced sentiment analysis pipelines fine-tune and hybridize BanglaBERT with lexicon/rule components \cite{mahmud2024enhancing}, or apply it for domain tasks like hyperpartisan news detection with semi-supervised learning and explainability \cite{hasan2025banglaBERT}. Alongside, general-purpose monolingual models for Bangla (e.g. “Bangla-Bert-Base” by Sagor Sarker et al.) have also been proposed and used across classification and NER tasks \cite{bangla-bert-github}.

Complementing these, newer model lines push toward generative and evaluation capacities in Bangla. TigerLLM, is a suite of Bangla LLMs trained on large Bangla corpora and shows gains over prior open and proprietary models across Bangla benchmarks \cite{raihan2025tigerllm}. In the programming domain, TigerCoder introduces dedicated Bangla code LLMs (1B and 9B) and the MBPP-Bangla benchmark, reporting 11 to 18 \% Pass@1 improvement over multilingual baselines \cite{tigercoder2025}. In evaluation, BenLLMEval  provides a wide evaluation of off-the-shelf LLMs (GPT-3.5, LLaMA-2, Claude, etc.) on Bangla tasks (summarization, QA, paraphrase, classification), revealing substantial performance gaps in zero-shot settings \cite{kabir2023benllmeval}. The BEnQA benchmark  offers parallel Bengali–English QA and reasoning tasks derived from exam questions; it shows that chain-of-thought prompting helps reasoning tasks and that including English context can improve performance in Bengali \cite{shafayat2024benqa}.

Despite advances in modeling, most existing works treat the language model as a single-step generator without built-in mechanisms for grounding, correction, or iteration. In broader NLP and code domains, however, robust generation systems increasingly incorporate retrieval-augmented architectures (e.g. RAG), cross-lingual retrieval for low-resource grounding, retrieval‐augmented data augmentation (RADA), multi-stage or hierarchical retrieval (e.g. for code), and iterative refinement via coder–reviewer loops or test-driven feedback. These techniques have been shown to reduce hallucination, improve factual grounding, and correct logical or syntactic errors in generated outputs.

These retrieval, review, and iteration techniques remain underexplored in Bangla and especially in Bangla–code generation. In this work, we explicitly address that gap by combining Bangla-focused models (e.g. TigerLLM, TigerCoder) with retrieval-based prompt augmentation, a separate reviewer module, and iterative self-refinement. This hybrid design aims to boost reliability and real-world usability in Bangla code generation systems.

\section{Experimental Setup}

All models were configured with the following default generation parameters: \texttt{temperature} = 0.7, \texttt{top\_p} = 0.9, and \texttt{max\_new\_tokens} = 1024. Each query generated \texttt{n} = 1 output sample per decoding pass.

\section{Model Prompts}
\label{sec:appendix_prompts}

This section details the prompts used in our \textbf{Bangla2Py} framework. The prompts are designed to guide the Large Language Models (LLMs) through the code generation, refinement, and review stages. Placeholders like \texttt{\{instruction\}} are dynamically populated by the pipeline.

\subsection{Coder Model Prompts}

\begin{figure*}[ht]
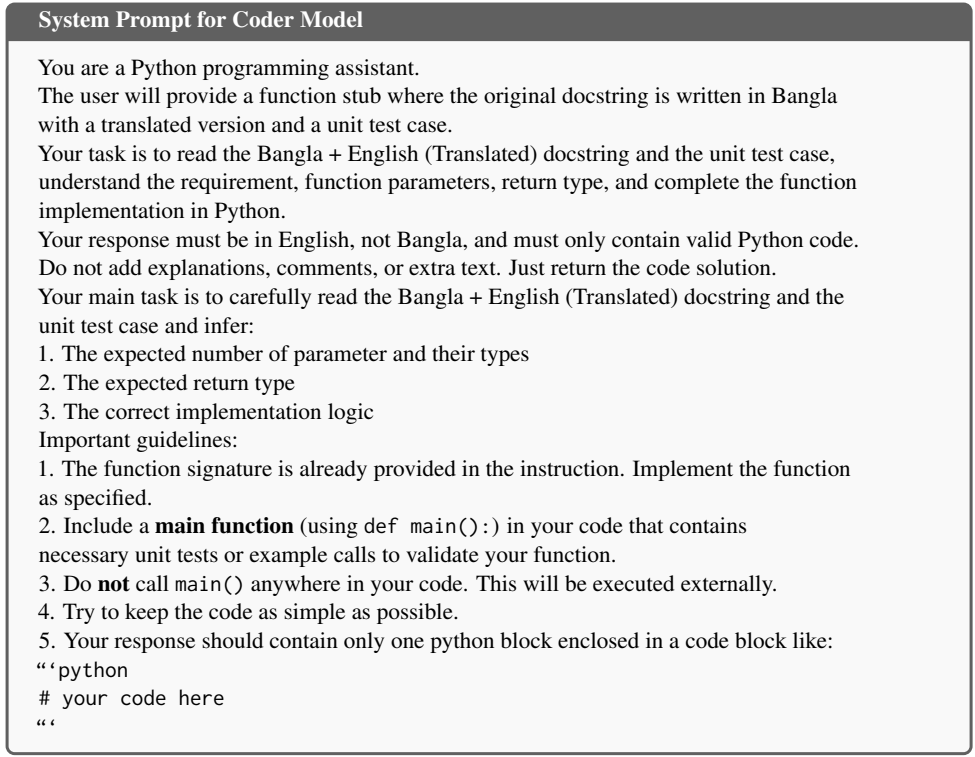

\centering
\resizebox{0.8\linewidth}{!}{%
\begin{tcolorbox}[
    title={System Prompt for Coder Model},
    colback=gray!5!white,
    colframe=gray!75!black,
    fonttitle=\bfseries,
    breakable,
    enhanced,
    listing only,
    listing options={basicstyle=\small\ttfamily,breaklines=true}
]
You are a Python programming assistant. 

The user will provide a function stub where the original docstring is written in Bangla\\
with a translated version and a unit test case.\\
Your task is to read the Bangla + English (Translated) docstring and the unit test case,\\
understand the requirement, function parameters, return type, and complete the function\\
implementation in Python.\\
Your response must be in English, not Bangla, and must only contain valid Python code.\\
Do not add explanations, comments, or extra text. Just return the code solution.\\
Your main task is to carefully read the Bangla + English (Translated) docstring and the\\
unit test case and infer:\\
1. The expected number of parameter and their types\\
2. The expected return type\\
3. The correct implementation logic

Important guidelines:\\
1. The function signature is already provided in the instruction. Implement the function\\
   as specified.\\
2. Include a \textbf{main function} (using \texttt{def main():}) in your code that contains\\
   necessary unit tests or example calls to validate your function.\\
3. Do \textbf{not} call \texttt{main()} anywhere in your code. This will be executed externally.\\
4. Try to keep the code as simple as possible.\\
5. Your response should contain only one python block enclosed in a code block like:\\
\texttt{```python}\\
\texttt{\# your code here}\\
\texttt{```}
\end{tcolorbox}
}
\caption{System prompt for the coder model.}
\label{lst:generator_system_prompt}
\end{figure*}

The \textbf{Coder LLM} is the first stage of our system and is responsible for writing the initial Python solution. It receives both the Bangla task description and its English translation, along with a set of retrieved examples and the provided unit tests.  
The coder's system prompt clearly defines its role as a Python code generator and instructs it to produce only executable code — no explanations or comments (Figure~\ref{lst:generator_system_prompt}).  
The main task prompt includes several few-shot examples followed by the current problem. Each example shows the task instruction (in both languages), the correct solution, and unit tests (Figure~\ref{lst:generator_main_template}).  
If the generated code fails any test, the coder receives a short feedback message describing the error type (e.g., syntax error, timeout, or assertion failure) along with a fix hint (Figure~\ref{lst:failed_attempt_template}). It then regenerates an improved version in the next iteration.  
This feedback-guided prompting helps the coder LLM progressively refine its output and produce cleaner, test-ready code with a built-in \texttt{main()} function for validation.

\begin{figure*}[ht]
\centering
\resizebox{0.8\linewidth}{!}{%
\begin{tcolorbox}[
    title={Main Prompt Template for Coder},
    colback=blue!5!white,
    colframe=blue!75!black,
    fonttitle=\bfseries,
    breakable,
    enhanced,
    listing only,
    listing options={basicstyle=\small\ttfamily,breaklines=true}
]
\{examples\}

>> Your Task\\
> Instruction\\
\texttt{```python}\\
\texttt{def \{function\_call\}:}\\
\texttt{~~~~"""\{instruction\}"""}\\
\texttt{~~~~"""Translated: \{instruction\_en\}"""}\\
\texttt{~~~~"""\{docstring\}"""}\\
\texttt{```}

Now complete the python code for the function '\texttt{\{function\_name\}}' and add a\\
'\texttt{main}' function with unit tests. You should use the '\texttt{check}' function for unit tests,\\
which is helpful for debugging. For example:

\texttt{```python}\\
\texttt{def \{function\_call\}:}\\
\texttt{~~~~\# Your code}\\
\\
\texttt{def check(test\_id, test\_val, expected):}\\
\texttt{~~~~assert test\_val == expected, f"Test \{test\_id\}: Expected \{expected\}, got \{test\_val\}"}\\
\\
\texttt{def main():}\\
\texttt{~~~~\{check\_example\}}\\
\texttt{~~~~\# Add more unit tests}\\
\texttt{```}
\end{tcolorbox}
}
\caption{Main prompt template for the coder, which includes few-shot examples and the current task.}
\label{lst:generator_main_template}
\end{figure*}

\begin{figure*}[ht]
\centering
\resizebox{0.8\linewidth}{!}{%
\begin{tcolorbox}[
    title={Failed Attempt Feedback Template},
    colback=red!5!white,
    colframe=red!75!black,
    fonttitle=\bfseries,
    breakable,
    enhanced,
    listing only,
    listing options={basicstyle=\small\ttfamily,breaklines=true}
]
>> Last failed code

> Response:\\
\texttt{\{last\_response\}}

> Error:\\
\texttt{\{last\_error\}}

> Suggested Fix:\\
\texttt{\{fix\_instructions\}}
\end{tcolorbox}
}
\caption{Template for providing feedback to the coder model after a failed execution attempt. This is appended to the main prompt during the self-refinement loop.}
\label{lst:failed_attempt_template}
\end{figure*}

\subsection{Reviewer Model Prompts}

The \textbf{Reviewer LLM} acts as the second stage and takes the code produced by the coder, along with the original Bangla–English instructions and all test cases (both given and generated).  
Its prompt defines the role of a “code reviewer” — focusing on improving correctness, readability, and coverage of edge cases without changing the function signature.  
The reviewer checks for logical mistakes, inefficient loops, missing validations, or weak test coverage. It then returns a refined version of the code, adds extra corner-case tests, and ensures the final version passes both visible and hidden cases.  
If errors are still detected, the reviewer can repeat this process with updated feedback until all tests are passed or a retry limit is reached.

\begin{figure*}[ht]
\centering
\resizebox{0.8\linewidth}{!}{%
\begin{tcolorbox}[
    title={System Prompt for Reviewer Model},
    colback=green!5!white,
    colframe=green!75!black,
    fonttitle=\bfseries,
    breakable,
    enhanced,
    listing only,
    listing options={basicstyle=\small\ttfamily,breaklines=true}
]
You are a Python code reviewer and programming assistant.

The user will provide a function stub or implementation where the original docstring\\
is written in Bangla with a translated English version, along with unit test cases.\\
Your task is to:\\
1. Do not alter the given function signature.\\
2. Review the implementation for correctness, clarity, efficiency, and robustness.\\
3. Refactor or improve the implementation if needed, but the function signature must\\
   remain identical.\\
4. Ensure the function works correctly not only for the provided tests but also for\\
   \textbf{hidden test cases} and \textbf{corner cases} (e.g., empty inputs, boundary values,\\
   invalid values, very large inputs).\\
5. Add a \texttt{main} function with unit tests that use the provided \texttt{check} function.\\
6. Include the given test cases and add additional edge/corner case tests that a\\
   hidden evaluator might check.\\
7. Do not add explanations, comments, or extra text. Just return the code solution.\\

Important guidelines:\\
1. The function signature is already provided. Implement or refactor the function\\
   as specified.\\
2. Include a \textbf{main function} (using \texttt{def main():}) that contains both the given\\
   unit tests and extra corner/hidden-case tests you find necessary.\\
3. Do \textbf{not} call \texttt{main()} anywhere in your code. It will be executed externally.\\
4. Keep the code clean, correct, and as simple as possible while ensuring it passes\\
   all tests, including edge and hidden cases.\\
5. Your response must be only one valid Python code block enclosed in triple backticks:\\
\texttt{```python}\\
\texttt{\# your code here}\\
\texttt{```}
\end{tcolorbox}
}
\caption{System prompt for the reviewer model.}
\label{lst:reviewer_system_prompt}
\end{figure*}

\begin{figure*}[ht]
\centering
\resizebox{0.8\linewidth}{!}{%
\begin{tcolorbox}[
    title={Main Prompt Template for Reviewer},
    colback=green!5!white,
    colframe=green!75!black,
    fonttitle=\bfseries,
    breakable,
    enhanced,
    listing only,
    listing options={basicstyle=\small\ttfamily,breaklines=true}
]
>> Your Task\\
The following function is already implemented:\\
~~~~\texttt{"""\{instruction\}"""}\\
~~~~\texttt{"""Translated: \{instruction\_en\}"""}\\
\texttt{```python}\\
~~~~\texttt{\{existing\_code\}}\\
\texttt{```}
\end{tcolorbox}
}
\caption{Main prompt template for the reviewer model.}
\label{lst:reviewer_main_template}
\end{figure*}

\subsection{Few-Shot Example Template}

To help both LLMs generalize better, we use retrieval-augmented few-shot examples in the prompts.  
The system retrieves the top-$k$ most similar problems from the bilingual database using both Bangla and English task texts. Each example includes:
\begin{itemize}
    \item The Bangla and English instructions,
    \item The reference Python solution, and
    \item The corresponding unit tests.
\end{itemize}
These examples are formatted in a consistent template and placed before the current task in the prompt (see Figure~\ref{lst:example_template}).  
This structure lets the models recognize patterns in how Bangla instructions map to Python logic, guiding them to produce correct and well-structured code even for unseen problems.

\begin{figure*}[ht]
\centering
\resizebox{0.8\linewidth}{!}{%
\begin{tcolorbox}[
    title={Few-shot Example Template},
    colback=orange!5!white,
    colframe=orange!75!black,
    fonttitle=\bfseries,
    breakable,
    enhanced,
    listing only,
    listing options={basicstyle=\small\ttfamily,breaklines=true}
]
>> Example \texttt{\{idx\}}:\\
> Instruction\\
\texttt{```python}\\
\texttt{def \{function\_call\}:}\\
\texttt{~~~~"""\{instruction\}"""}\\
\texttt{~~~~"""Translated: \{instruction\_en\}"""}\\
\texttt{~~~~"""\{docstring\}"""}\\
\texttt{```}\\
> Solution\\
\texttt{```python}\\
\texttt{\{solution\}}\\
\\
\texttt{def check(test\_id, test\_val, expected):}\\
\texttt{~~~~assert test\_val == expected, f"Test \{test\_id\}: Expected \{expected\}, got \{test\_val\}"}\\
\\
\texttt{def main():}\\
\texttt{~~~~\{test\_main\}}\\
\texttt{```}
\end{tcolorbox}
}
\caption{Template for formatting each of the k-nearest examples for retrieval-augmented generation.}
\label{lst:example_template}
\end{figure*}

\begin{figure*}[h]
\centering
\resizebox{0.8\linewidth}{!}{%
\begin{tcolorbox}[
    title={System Prompt for Translator Model},
    colback=gray!5!white,
    colframe=gray!75!black,
    fonttitle=\bfseries,
    breakable,
    enhanced,
    listing only,
    listing options={basicstyle=\small\ttfamily,breaklines=true}
]
Translate the following Bangla Python Code Instruction to English and only return the\\
English translation. Do not change the example function and parameter names and only\\
update the function parameter types and return variable types of Example function\\
prototype to actual python syntax based on the provided unit test. Do not give the full\\
code implementation. Just give the updated prototype.\\

Use the following glossary for translation: \{glossary\}\\

Unit Test: \{test\}
\end{tcolorbox}
}
\caption{System prompt for the translator model.}
\label{lst:translator_system_prompt}
\end{figure*}

\begin{figure*}
    \centering
    \includegraphics[width=1\linewidth]{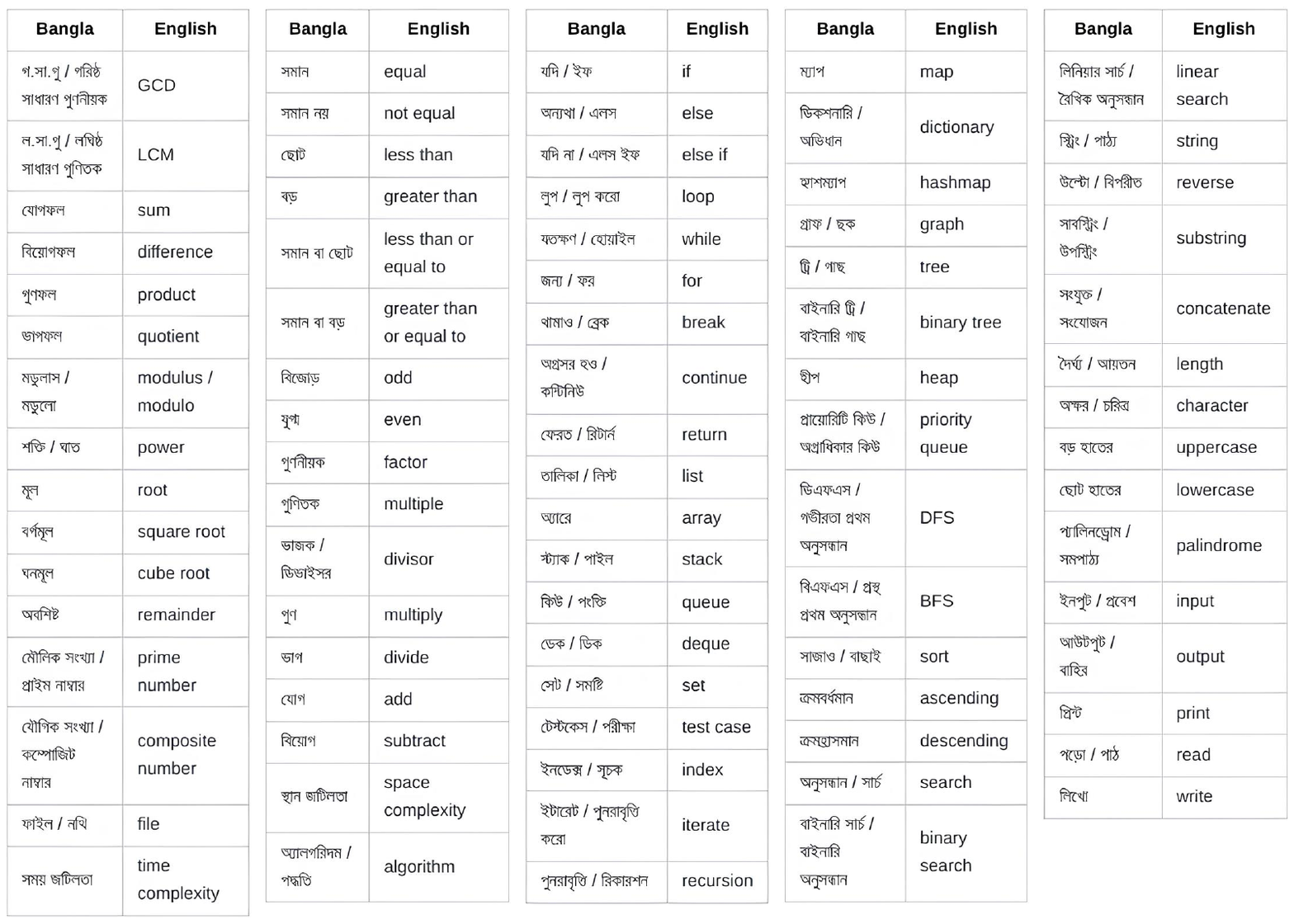}
    \caption{Glossary for the translation prompt}
    \label{fig:glossary}
\end{figure*}

\section{Error Refinement}
The iterative feedback follows an augmentation protocol as outlined in Table~\ref{tab:error_hints}.

\begin{table}[h]
\centering
\small
\setlength{\tabcolsep}{3pt}
\renewcommand{\arraystretch}{1.05}
\begin{tabular}{@{}|p{2.2cm}|p{5cm}|@{}}
\hline
\makecell{\textbf{Error Type}} & 
\makecell{\textbf{Feedback Hint / Guidance}} \\ \hline

Syntax Error &
Check indentation, missing colons, or parentheses; ensure valid Python syntax. \\ \hline

Runtime Error &
Ensure variables are initialized and referenced correctly; verify data types and control flow. \\ \hline

Assertion Failure &
Compare expected vs.\ actual outputs; review logical steps and boundary conditions. \\ \hline

Timeout Error &
Optimize loops or recursion; include clear termination conditions. \\ \hline

System Exit &
Avoid abrupt exits; allow the program to complete execution normally. \\ \hline

\end{tabular}
\caption{Error categories and corresponding feedback hints used in prompt augmentation.}
\label{tab:error_hints}
\end{table}

\section{Ablation Study}

To analyze the contribution of each component in \textbf{\tool}, we perform ablation experiments using the \textbf{Lg Exaone Deep 32B} model on the BLP-2025 development set. The best full configuration achieves a Pass@1 accuracy of \textbf{95.5\%}, and all reported variations are measured relative to this setting. Each ablation disables or modifies a single module while keeping the rest of the pipeline fixed.

\subsection{Effect of English Translation}

We first evaluate the role of bilingual translation. When the system relies solely on Bangla instructions without their English counterparts, comprehension drops significantly. The LLM often fails to parse algorithmic phrases and control keywords written in Bangla. As shown in Table~\ref{tab:ablation-translation}, removing the translation stage reduces Pass@1 accuracy by nearly 22 percent, confirming that current models still struggle to reason directly over Bangla-only text.

\begin{table}[h]
\centering
\small
\begin{tabular}{@{}lc@{}}
\toprule
\textbf{Setting} & \textbf{Pass@1 (\%)} \\ \midrule
Full Model (Bangla + English) & 95.5 \\
Bangla Only & 73.6 \\ \bottomrule
\end{tabular}
\caption{Effect of English translation on Pass@1 accuracy (Lg Exaone Deep 32B, Dev Set).}
\label{tab:ablation-translation}
\end{table}

\subsection{Effect of Glossary-based Translation}

We also analyze the impact of the controlled translation glossary used for mathematical and algorithmic terms. 
Without this glossary, the translation model often produces inconsistent or incorrect terminology, confusing the Coder during reasoning.  
As shown in Table~\ref{tab:ablation-glossary}, removing the glossary results in a notable performance drop of over 7 points, confirming that LLMs struggle to translate some Bangla words properly, leading to incorrect function generation.

\begin{table}[h]
\centering
\small
\begin{tabular}{@{}lc@{}}
\toprule
\textbf{Setting} & \textbf{Pass@1 (\%)} \\ \midrule
With Glossary (Full Model) & 95.5 \\
Without Glossary & 88.2 \\ \bottomrule
\end{tabular}
\caption{Effect of using the controlled translation glossary.}
\label{tab:ablation-glossary}
\end{table}

\subsection{Effect of Feedback Loop}

Next, we disable the iterative self-refinement mechanism. Without execution feedback or re-prompting, the model cannot correct runtime or logic errors, leading to a steep performance drop. Table~\ref{tab:ablation-feedback} shows that accuracy declines by more than 25 percent, emphasizing that feedback-driven correction is vital for reliable synthesis.

\begin{table}[h]
\centering
\small
\begin{tabular}{@{}lc@{}}
\toprule
\textbf{Setting} & \textbf{Pass@1 (\%)} \\ \midrule
Full Model & 95.5 \\
Without Feedback Loop & 69.8 \\ \bottomrule
\end{tabular}
\caption{Impact of feedback-driven refinement.}
\label{tab:ablation-feedback}
\end{table}

\subsection{Effect of Reviewer LLM}

To measure the Reviewer’s contribution, we bypass the second-stage review and directly execute the Coder output. Although the generated code remains mostly functional, it lacks stylistic polish and robustness on edge cases. Table~\ref{tab:ablation-reviewer} shows a moderate decline of about 5 percent, verifying that the Reviewer mainly improves coverage and reliability.

\begin{table}[h]
\centering
\small
\begin{tabular}{@{}lc@{}}
\toprule
\textbf{Setting} & \textbf{Pass@1 (\%)} \\ \midrule
Full Model & 95.5 \\
Without Reviewer & 90.4 \\ \bottomrule
\end{tabular}
\caption{Effect of disabling the Reviewer LLM.}
\label{tab:ablation-reviewer}
\end{table}

\subsection{Number of Feedback Iterations}

We vary the maximum feedback iterations ($M$) to observe convergence behavior. As shown in Table~\ref{tab:ablation-iterations}, fewer iterations significantly reduce success rate since many tasks require multiple refinement cycles. Beyond five iterations, improvements saturate.

\begin{table}[h]
\centering
\small
\begin{tabular}{@{}lc@{}}
\toprule
\textbf{Max Iterations ($M$)} & \textbf{Pass@1 (\%)} \\ \midrule
1 & 84.1 \\
3 & 92.4 \\
5 & 95.5 \\
7 & 95.5 \\ \bottomrule
\end{tabular}
\caption{Effect of limiting feedback iterations ($M$).}
\label{tab:ablation-iterations}
\end{table}

\subsection{Effect of Retrieval Augmentation (RAG)}

We compare our retrieval-augmented setup against a manually few-shot configuration.  
In the manual setup, the examples are fixed and not selected dynamically based on similarity, while the RAG variant retrieves the top-$k$ relevant bilingual examples for each new task.  
As Table~\ref{tab:ablation-rag} shows, retrieval augmentation provides a small but consistent improvement of about 1.3 points, indicating that example relevance matters more than sheer quantity.

\begin{table}[h]
\centering
\small
\begin{tabular}{@{}lc@{}}
\toprule
\textbf{Setting} & \textbf{Pass@1 (\%)} \\ \midrule
With RAG (Full Model) & 95.5 \\
Manual Few-shot (Fixed Examples) & 94.2 \\ \bottomrule
\end{tabular}
\caption{Comparison between manual few-shot and RAG-based prompting.}
\label{tab:ablation-rag}
\end{table}

\subsection{Number of Retrieved Examples ($k$)}

Finally, we study the impact of the retrieval context size. As Table~\ref{tab:ablation-k} shows, removing examples ($k=0$) severely hampers the model’s grounding ability, dropping performance below 70\%. Accuracy improves steadily up to $k=5$, after which marginal gains diminish due to context saturation.

\begin{table}[h]
\centering
\small
\begin{tabular}{@{}lc@{}}
\toprule
\textbf{Number of Examples ($k$)} & \textbf{Pass@1 (\%)} \\ \midrule
0 (No Examples) & 69.3 \\
3 & 88.9 \\
5 (Full) & 95.5 \\
7 & 94.7 \\ \bottomrule
\end{tabular}
\caption{Effect of retrieved example count ($k$).}
\label{tab:ablation-k}
\end{table}

\subsection{Comprehensive Summary}

Table~\ref{tab:ablation} consolidates all variants. The results confirm that English translation and the feedback loop contribute the largest performance boosts, while the glossary, reviewer, and RAG components further improve consistency, code quality, and generalization.

\begin{table*}[h]
\centering
\small
\begin{tabular}{@{}lcccccc@{}}
\toprule
\textbf{Configuration} & \textbf{Translation} & \textbf{Glossary} & \textbf{Feedback Loop} & \textbf{Reviewer} & \textbf{RAG} & \textbf{Pass@1 (\%)} \\ \midrule
Full \tool Pipeline & Yes & Yes & Yes & Yes & Yes & \textbf{95.5} \\
Without Translation & No & Yes & Yes & Yes & Yes & 73.6 \\
Without Glossary & Yes & No & Yes & Yes & Yes & 88.2 \\
Without Feedback Loop & Yes & Yes & No & Yes & Yes & 69.8 \\
Without Reviewer & Yes & Yes & Yes & No & Yes & 90.4 \\
Manual Few-shot (No RAG) & Yes & Yes & Yes & Yes & No & 94.2 \\
Fewer Iterations ($M$ = 1) & Yes & Yes & Yes & Yes & Yes & 84.1 \\
Fewer Examples ($k$ = 3) & Yes & Yes & Yes & Yes & Yes & 88.9 \\
No Examples ($k$ = 0) & Yes & Yes & Yes & Yes & Yes & 69.3 \\ \bottomrule
\end{tabular}
\caption{Comprehensive ablation results on the BLP-2025 development set using Lg Exaone Deep 32B.}
\label{tab:ablation}
\end{table*}

\section{Algorithm}\label{sec:appendix_algo}
Algorithm \ref{alg:banglaforge} shows the pseudocode of our pipeline.

\begin{algorithm*}[t]
\caption{Algorithm of \tool}
\label{alg:banglaforge}
\begin{algorithmic}[1]
\State \textbf{Input:} BanglaInstruction $P_b$, PublicUnitTests $T$
\State \textbf{Output:} ExecutableCode
\State $M \gets$ maximum retry limit
\State $attempt \gets 0$
\State EnglishInstruction, $P_e$ $\gets$ TranslatorLLM.translate($P_b$)
\State Examples, $E$ $\gets$ Database.retrieveExamples($P_b$, $P_e$)

\State PromptCoder $\gets$ constructPrompt($P_b$, $P_e$, $T$, $E$)

\While{$attempt < M$}
    \State $attempt \gets attempt + 1$
    \State ($c$, $T_c$) $\gets$ CoderLLM.generate(PromptCoder)
    \State PromptReviewer $\gets$ constructReviewPrompt($c$, $T_c$)
    \State ($c_r$,$T_r$) $\gets$ ReviewerLLM.refine(PromptReviewer)

    \State Result $\gets$ executeCode($c_r$, $T$ $\cup$ $T_c$ $\cup$ $T_r$)

    \If{Result.allTestsPassed}
        \State \textbf{return} $c_r$
    \Else
        \State Feedback $\gets$ generateFeedback(Result.errors)
        \State PromptCoder $\gets$ updatePromptWithFeedback(PromptCoder, Feedback)
    \EndIf
\EndWhile

\end{algorithmic}
\end{algorithm*}

\section{Failure Cases and Dataset Limitations}

The dataset for Bangla-to-Python code generation was created by translating existing English datasets MBPP (Mostly Basic Python Problems) and HumanEval into Bangla using machine translation. While this approach enables rapid dataset construction, it introduces several limitations that affect both dataset quality and model performance.

\subsection{Semantic and Syntactic Translation Errors}
Machine translation occasionally produces Bangla sentences that are grammatically incorrect or semantically ambiguous. Such translations may hinder a model’s ability to correctly interpret the input and generate the intended Python code.  
For example:
\begin{itemize}
    \item The English adjective ``even'' was translated as \raisebox{-1ex}{\includegraphics[width=0.19\linewidth]{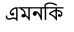}} instead of the more contextually accurate \raisebox{-1ex}{\includegraphics[width=0.16\linewidth]{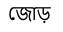}} in cases where \textit{even} refers to parity in numbers. This leads to semantic confusion and misinterpretation of the question context.
\end{itemize}

\subsection{Incorrect or Misleading Terminology for Programming Concepts}
Programming terms often lack direct equivalents in Bangla. Machine translation systems attempt to generate literal translations, but these often fail to capture technical meaning.  
For example:
\begin{itemize}
    \item The English term \textit{``Map''} was translated to \raisebox{-1ex}{\includegraphics[width=0.19\linewidth]{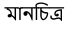}} (meaning a geographic map in Bangla), instead of referring to \textit{Map} as in a data structure such as HashMap or dictionary. This causes ambiguity, making it challenging for both humans and models to interpret correctly.
    \item Similarly, terms like \textit{stack}, \textit{queue}, \textit{hashmap}, or \textit{dictionary} may be incorrectly translated, or not translated at all, resulting in inconsistent terminology across the dataset.
\end{itemize}

\subsection{Loss of Context or Intent}
Machine translation may fail to preserve the precise context or intent of the original English instructions. Programming problems often rely on subtle nuances, and even small changes in wording can alter the meaning of a problem. This issue is exacerbated when the translated text uses uncommon or unnatural phrasing, reducing clarity for model training.

\subsection{Lack of Standardized Technical Vocabulary}
Bangla currently lacks standardized technical vocabulary for many programming concepts, leading to inconsistent translations. In some cases, the same English term is translated differently across dataset entries. This inconsistency makes it difficult for a model to reliably learn the intended mapping from Bangla instructions to Python code.

\subsection{Impact on Model Performance}
These translation-related issues contribute to notable failure cases in Bangla-to-Python code generation. Models trained on such data may misinterpret problem statements, produce incorrect code, or fail to generalize to unseen examples. Addressing these limitations would require:
\begin{itemize}
    \item Careful human curation of translations for correctness and consistency.
    \item Development of a standardized Bangla programming lexicon.
    \item Use of bilingual glossaries to retain original technical terms where necessary.
\end{itemize}


\end{document}